\def\im{\hbox{Im}\,}
\def\re{\hbox{Re}\,}

\documentclass[12pt]{article}
\usepackage{hyperref}
\usepackage{amsfonts}

\title{Nonlinear Quantum Mechanics at the Planck Scale}
\author{George Svetlichny\footnote{Departamento de Matem\'atica, Pontif\'{\i}cia Universidade Cat\'olica, Rio de Janeiro, Brazil \newline
svetlich@mat.puc-rio.br \hfill \url{http://www.mat.puc-rio.br/\~svetlich}}}
\begin{document}
\maketitle
\begin{abstract}
Linearity of quantum mechanics is an emergent feature at the Planck scale, along  with the manifold structure of space-time. In this regime the usual causality violation objections to nonlinearity do not apply, and  nonlinear effects can be of comparable  magnitude to the linear ones and still be highly suppressed at low energies.  This can offer alternative approaches to quantum gravity and to the evolution of the early universe.
\end{abstract}
{\obeylines
}
\section{Introduction}

I shall adopt what would be a physicist's conception of nonlinear quantum mechanics. It would be a theory that (1) at low energies reduces to standard linear quantum mechanics, and (2) involves, in an essential way, nonlinear operators in lieu of linear ones. This is not the view adopted in a large part of the literature dedicated to the subject. Investigations of abstract structures may ignore the first point if no immediate confrontation with reality is  contemplated, and the second point may only be implicit if an operator approach is not used. At a meeting of quantum logicians such as this IQSA 2004, anyone who considers quantum logics not representable in Hilbert space may be though to be essentially dealing with nonlinear quantum mechanics. For my purposes though the stated view is essential. One should quickly point out though that there is no such theory, at best one has only a few exploratory results.

It may be useful to begin with some historical and bibliographical perspectives and try to answer the {\em who?\/}, {\em what?\/}, {\em how\/}? and {\em why\/}? of the field. Nonlinear quantum mechanics does not have a large literature and its content is very varied. 
A rough survey in arXiv reveals roughly 119 articles by 95 authors since 1991. I posted a list of these on the archives (Svetlichny 2004c). In this compilation I only included articles that deal with supposed fundamental nonlinearities in the quantum formalism  and left out those that deal with nonlinearities but do not question the fundamental linearity of the underlying quantum physics.

The above survey of course leaves out contributions from the pre-internet era and some important references such as  Bialynicki-Birula and Mycielski 1976,  Bugajski 1991, Gisin 1984, 1989, 1990, Haag and Bannier 1978, Kibble 1979, Kostin 1972, Polchinski 1991, and Weinberg 1989a, 1989b.

From this one can guess that the total literature contains at most a few hundred works by roughly the same number of authors. Thus few authors contribute regularly, most contribute once or sporadically.  This is already an unusual situation for a topic that has been around for over twenty years.

What is also unusual are some of the things {\em said\/} about nonlinear quantum mechanics, for instance:

\begin{enumerate}
\item It is essentially classical:  Bugajski 1991, Haag~and~Bannier 1978.
\item It allows communication between Everett histories: Polchinski 1991.
\item It makes experiments react to the content of the experimenter's mind: Polchinski 1991.
\item It violates causality:  Polchinski 1991; Gisin 1984, 1989, 1990; Svetlichny 1998; L\"ucke 1999.
\item It solves the ``measurement problem":  Hansson 2000.
\item Solves NP-complete and \#P problems in polynomial time: Abrams and Lloyd 1998.
\item May violate space-time symmetries:  Parwani 2004, Svetlichny 1995. 

\item It is necessary for introspection:  H\"ubsch 1998.
\item It is involved in black-hole dynamics: Yurtserver and Hockney 2004. 
\item It is necessary for homogeneous quantum cellular automata on Euclidean space: Meyer 1996.
\item It is necessary in the presence of closed time-like curves: Cassidy 1995.
\end{enumerate}

Now I shall not discuss the merit of any of these claims, but only call attention to the great variety of some very fundamental scientific issues that nonlinear quantum mechanics forces us to consider and reevaluate.  Why should this be so? My tentative answer to this is that the observed quantum linearity is related to space-time structure, and space-time obviously bears upon all our fundamental concerns. My view is that space-time and quantum mechanics is a unified whole and one cannot understand the  one without the other. Both are emergent aspects of a more fundamental theory. When one is dealing on the level of such a theory, the emergent quantum mechanics may very well be nonlinear and linearity comes about because it must eventually act in a space-time arena. Let us then form the hypothesis that  a nonlinear quantum mechanics, very close to the linear one, is the true theory at the level of emergence. How can one come to know anything about it? 

\section{Ways toward nonlinear Quantum Mechanics}

There are two ways leading to nonlinear quantum mechanics.

\begin{enumerate}
\item {\bf Willing.} Here nonlinearity is posited from the beginning. This is due either to  intelectual speculation or  axiomatics of more general mechanics.  Most proposals fall into this category.
\item {\bf Unwilling.} One is forced to consider nonlinearities in ostensively linear contexts. Nonlinearity comes as a surprise. Examples of such situations are:
\begin{enumerate}
\item Representations of current algebras: Doebner~and~Goldin 1992;
\item Dynamics of D0 branes in non-critical string theory: Mavromatos~and~Szabo 2001;
\item Introspective quantum mechanics: H\"ubsch 1998;
\item Quantum evolution in acausal space-times: Cassidy 1995;
\item Quantum cellular automata on Euclidean lattices:  Meyer 1996. 
\end{enumerate}
\end{enumerate}
The degree of ``surprise" in these situations of of course subjective and my personal ``unwilling" list is unstable, but the first item seems to be always firmly in place.

One of the motivations for the willing approach to nonlinearity is solution of some fundamental problem of contemporary science. A sample of these is:  (1) in quantum gravity:  time, probability, black hole information, classicality, etc.\ (2)in cosmology: horizon, flatness, entropy, defects, dark matter, dark energy, coincidence etc.\ (3)
in quantum mechanics:  measurement, decoherence etc.\ (4) in general: computability, consciousness, etc.
 
Now since nonlinear quantum mechanics calls for a  broadband modification of all our fundamental physical theories, one can expect that {\em any\/} nonlinear quantum mechanical theory will {\em  appear\/} to solve some of these fundamental problems, that is, will address some of these problems better than the existing theories.  This means that a resolution of one or other of these problems by a nonlinear theory cannot be considered a strong reason for its adoption.

Based on the above considerations I have adopted the following guiding rules for trying to approach the hypothetical true nonlinear theory. 
 
\begin{enumerate}
\item The unwilling nonlinearities are more likely to be closer to the true theory than the willing ones.
\item Widespread properties of studied nonlinearities are more likely to be true of the true theory.
\item One should not be motivated by the desire to solve any particular ``fundamental problem". 
\end{enumerate}
Based on this I here  will focus on the (unwilling) Doebner-Goldin nonlinearities (Doebner and Goldin 1992) and address the  (widespread) causality issue  (Gisin  1984, 1989, 1990; Luecke 1991, Polchinski 1991; Svetlichny 1998).
\section{The Doebner-Goldin nonlinearity}

Doebner and Goldin (1992) studied representations of non-relativistic current algebras, which in particular involves unitary representations of the diffeomorphism group of ordinary Euclidian space \({\mathbb R}^n\). One such notable representation has a non-trivial cocycle and is given in \({\cal H}=L^2({\mathbb R}^n)\) by 
\[(V(\phi)\Psi)(x)=\exp[iD\ln {\cal J}_\phi(x)]\Psi(\phi(x))[{\cal J}_\phi(x)]^{1/2}\]
where \(\Psi\in {\cal H}\), \(\phi:{\mathbb R}^n\to {\mathbb R}^n\) is a diffeomorphism, \({\cal J}_\phi(x)\) is its jacobian  and \(D\) is a physical constant.

From this representation one can construct the density \(\rho\) and current  \({\bf J}\) operators of a non-relativistic quantum theory. In contrast to representations with a trivial cocycle, these density and current operators do not satisfy a continuity equation but instead a  Fokker-Planck equation:
\[\partial_t\rho=-\nabla \cdot {\bf J}+D\nabla^2\rho.\]
No linear quantum system is consistent with this, but nonlinear ones are, the simplest given by the Doebner-Goldin  equation
\begin{equation}\label{dg}
i\hbar\partial_t\psi =-\frac{\hbar^2}{2m}\Delta \psi +
iD\hbar\left(\Delta\psi +
\frac{|\nabla\psi|^2}{|\psi|^2}\psi\right).
\end{equation}

One can add to the right-hand side any term of the form \(R(\psi)\psi\). Where \(R\) is any {\em real\/} not necessarily linear operator applied to \(\psi\) and which is homogeneous of degree zero, that is \(R(z\psi)=R(\psi)\).

Now representations of the diffeomorphism group is certainly a highly respectable mathematical topic. That nonlinear quantum systems are somehow connected to them is probably the strongest reason to give them further thought, especially since diffeomorphism related issues are germane to physics at the Planck level.

\section{The separation property}
 
 One important property that a nonlinear evolution can satisfy is that of {\em  separability\/} which is a nonlinear generalization of lack of interaction. Understanding non-interacting systems well is an important step toward understanding the interacting ones.  Assuming that states can still be represented by wave functions, the separation property is: 
\[
E_s(t_2,t_1)(\Psi_1\otimes\Psi_2)= 
 E_{s_1}(t_2,t_1)(\Psi_1) \otimes E_{s_2}(t_2,t_1)(\Psi_2).
\]
Here \(E\) is the evolution operator and the \(s_i\) the species indicator of  \(n_i\) particles. All particles belong to different species. What this equation states is that non-correlated systems continue non-correlated and 
is a nonlinear generalization of lack of interaction

The evolution is governed by a not necessarily linear Shr\"odinger equation
\[i\hbar\partial_t\Psi_s=H_s\Psi_s.\]

The  separation property necessarily implies (Goldin and Svetlichny 1994) that
\[H_s\Psi=K_s\Psi+p\ln|\Psi_s|\,\Psi_s+iq\ln(\Psi_s/\bar\Psi_s)\,\Psi_s\]
where \(p\,\ln|\Psi_s|\,\Psi_s\) is the Bialynicki-Birula and Mycielski (19976) term,  \linebreak \(iq\ln(\Psi_s/\bar\Psi_s)\,\Psi_s\) is the Kostin (1972) term, \(p\) and \(q\) are universal physical constants, and \(K_s\) is  homogeneous
\[K_s(z\Psi)=zK_s(\Psi).\]
 
A two particle Schr\"odinger operator is build-up from  one particle operators by
\begin{equation}\label{tpse}
K_{ab}\Psi=K^{(1)}_a\Psi+K^{(2)}_b\Psi+Q \Psi
\end{equation}
where \(K^{(i)}_s\)  is a one-particle operator acting on the \(i\)-th variable of \(\Psi\) and \(Q\) is an operator that vanishes identically on product functions.
This generalizes to an \(n\)-particle operator construction and one can introduce true \(n\)-particle effects that don't exist for smaller number of particles

The case for identical particles is more subtle. There are no nonlinear separating hierarchies of differential  Schr\"odinger operators (Svetlichny 2004b) and this constitutes another indication that linearity has something to do with space-time structure. 
 
\section{(Non)linearity and space-time}

In my view the following results point to a connection between linearity of quantum mechanics and space-time structure:
\begin{enumerate}
\item  Nonlinearity along with instantaneous state-collapse violates causality (enables superluminal signals): Polchinski 1991, Gisin 1989, L\"ucke 1999,
\item Causality implies linearity: Svetlichny 1998, Simon,~Bu\v{z}ek~and~Gisin 2001; 
\item Nonlinearity and internal symmetries imply new effects at each particle number: Svetlichny 1995;
\item Piron's covering law in quantum logic is connected to Lorentz covariance: Svetlichny 2000;
\item Differential separating equation for identical particles are linear: \linebreak Svetlichny 2004b. 
\end{enumerate}
 
This is a personal list, others who have come to similar conclusions would probably cite other sources. The results that particularly formed my viewpoint are my own in reference (Svetlichny 2000).  Inspired by the works of  W. Guz (1979, 1980) on the covering law and  R. Haag (1992) on local quantum theory this is an attempt to deduce the covering law (considered a close relative of linearity) through a local relativistic quantum logic. The covering law can be deduced from
\begin{enumerate}
\item  Lorentz covariance;
\item  Causality, that is, propositions belonging to space-like separated regions commute;
\item State-collapse; 
\item An abundance of space-like separated entangled states to be able to ``prepare at a distance" any given state (true of local relativistic quantum mechanics).
\end{enumerate}
The first interesting fact  about this is that if one insists on eliminating non-local state-collapse, for example through a version of the coherent histories approach (such as attempted in Svetlichny 1997), one cannot complete the deduction. This makes quantum mechanics  understandable only if one combines relativity and causality with some form of non-locality. Now, only a quantum space-time is capable of bridging the time-like and the space-like. The second fact is that if the argument shows universality of quantum mechanics, it must also apply to space-time related measurements. However, the argument assumes a classical Minkowski space-time, hence there is a fundamental inconsistency in the above approach. The conclusion is inescapable: {\em only {\em quantum\/} space-time can make quantum mechanics intelligible\/}.

Hence I come to my main conjecture, which was also voiced by other authors: 
\vskip 4pt
\noindent{\em Linear quantum mechanics is an emergent feature of ``quantum gravity" which may very well be nonlinear\/}. 
\vskip 4pt
See Markopoulou and  Smolin 2003, Parwani 2004,  Singh 2003, and Svetlichny 2004a, 2004b. 
 
\section{Nonlinear quantum mechanics and causality}
As was mentioned above, the appearance of superluminal signals seems to be a generic feature of nonlinear quantum theories. There have been several proposals for 
circumventing this apparent violation of causality (the following list is undoubtedly incomplete):
 \begin{enumerate}
\item The introduction of elementary mixtures, that is, ontological irreducibility of mixed states: Bona 1999, Czachor 1999, Gheorghiu-Svirschevski 2002.
\item Absence of self subsisting physical states in a nonlinear version of coherent histories:  Svetlichny 1997.
\item Modified measurement process: Kent 2002.
\end{enumerate}
As valiant as these efforts might be, one still cannot say that we have an explicit and consistent causal relativistic nonlinear quantum mechanical theory, however, {\em if nonlinearities are of Plank scale should one worry\/}?  

At Planck  energies space-time is thought to be ill defined, the causal structure also ill defined, and so it makes little sense to talk of its violation.
Lorentz invariance itself may be  broken  (Amelino-Camelia 2003), a hypothesis put forth to explain some cosmic ray phenomena (see Section \ref{nlqmecr}), which further casts doubt on the ultimate seriousness of the causality violation issues. In the end all the space-time difficulties of nonlinear quantum mechanics may not be  pernicious. It would take Planck energies to exhibit the effects, but then space-time itself becomes quantum and the apparent problems could have no problematic low energy consequences. The presence of such effects at the Planck scale could however completely transform our understanding of quantum space-time.
\section{Experimental situation}\label{nlqmecr} 
A series of experiments designed to test nonlinear effects of the Weinberg (1989a, 1989b) type show that these are about twenty orders of magnitude smaller than linear ones (Bollinger et al. 1989, Walsworth et al. 1990, Chupp and Hoare 1990, Majumder et al. 1990, Benatti and Floreanini 1996, 1999). While this is consistent with the hypothesis that such putative effects would only appear  on the Planck scale, one is inevitably led to ask, if so,  how can one become aware of them? Now there is at least one (possible) physical phenomenon for which nonlinear quantum mechanics is a ready-made explanation.  This occurs in cosmic ray physics (Svetlichny 2004a and references therein). 

Cosmic rays can scatter off the cosmic microwave background, and the cross section for scattering increasing with energy. As 
 there are no know nearby sources of such rays, one should not see any above a certain energy (the so called GZK cutoff). Apparently about twenty  such events have been seen, and though the existence of this effect is still being debated, 
 speculations abound concerning new physics that would explain them, such as quantum gravity, non-commutative space-time, lorentz symmetry breaking, etc.
 
All such explanations propose a modified dispersion relation instead of the usual \(E^2=m^2c^4+p^2\), typically: 
 \begin{equation}\label{moddr}
E^2=m^2c^4+p^2+\kappa\ell_pp^3
\end{equation}
where \(\kappa\) is of order unity and \(\ell_p\) the Planck length \((\hbar g/c^3)^{1/2}\approx 10^{-33}{\rm\ cm}\).
 
The highest cosmic ray energy seen is about \(10^{20.5}\,\hbox{ev.}\) 
Now Planck energy is about \({10^{28}\,\hbox{ev}}\) and so the de Broglie wavelength of such a particle is 
\[10^{28}/10^{20.5}\approx 3.17\times 10^7\,\hbox{Planck lengths}. \]
Though thirty million may seem large, it is small enough that deviations from a smooth manifold structure can already influence the propagation of the particle. Nature thus supplies us in our midst with true quantum gravity experiments, and we have access to them.

If one did not already have some beginnings of quantum gravity and non-commutative space-time theories, what would be a  reaction to being forced to use (\ref{moddr}) to explain the phenomenon? One could say that for some reason Lorentz covariance is broken, one could say that at high energies particle propagation uses higher order (greater than $2$) differential equations, or one could say quantum mechanics is nonlinear.  
In my view nonlinear quantum mechanics is the simplest {\em  prima-facie\/} explanation for the modified dispersion relation. It fits the working definition of nonlinear quantum mechanics given at the beginning of this paper provided one can argue that physics at this energy should already be described by second order differential equations.  No linear second order equation can produce a modified dispersion relation.
\section{Planck-scale nonlinear quantum effects}
 
If nonlinear quantum effects exist at Planck energies, how large can they be and still be consistent with the large experimental suppression at low energies? This is a hard question to get a handle on given a lack of high-energy nonlinear theories but one may get a hint looking at non-local signaling due to separated measurements. Now at the Planck scale, say in the early universe, there are no observers measuring things, however there may be decoherence effects having similar consequences (Hansson 2000). This makes measurement-related arguments relevant.   

Consider a not necessarily linear evolution operator \(E_t\)  and two conventional quantum observables \(A\) and \(B\). Assume \([A,B]=0\). The expected
value of \(B\) in the mixture resulting from a measurement of \(A\) on a state \(\Phi\)
followed by evolution for time \(t\) is
\[
{\cal E}(B,t|A)=
\sum_\lambda||P_\lambda\Phi||^2(E_t\phi^A_\lambda,BE_t\phi^A_\lambda)
\]
where \(P_\lambda\) is a spectral projection of \(A\) and \(\phi^A_\lambda\) is the eigenvector (assume for simplicity a non-degenerate spectrum). 
Consider now the difference when one introduces an alternative measurement \(A'\), again with \([A',B]=0\):
\[
\Delta(B,t|A,A')={\cal E}(B,t|A)-{\cal E}(B,t|A').
\]
Assume evolution is  described through a
 Schr\"odinger-type equation:
\[
i\hbar\partial_t\Psi = H\Psi
\]
for some generally nonlinear operator \(H\), and that the evolution is norm preserving, which is  equivalent to \(\im(\Phi,H\Phi)=0\). 
Expanding into a Taylor series one has:
\begin{eqnarray*}
{\cal E}(B,t|A) &=& {\cal E}(B,0|A)+ t{\cal E}_1(B|A)+ O(t^2) \\
\Delta(B,t|A,A') &=& \Delta(B,0|A,A')+ t\Delta_1(B|A,A')+ O(t^2).
\end{eqnarray*}
Now  \(\Delta(B,0|A,A')=0\) by the linear quantum mechanical no signal theorem and since \(B\) is
hermitian one has
\[
{\cal E}_1(B|A)=\frac{2}{\hbar}
\sum_\lambda||p_\lambda\phi||^2\im(B\phi^A_\lambda,H\phi^A_\lambda).
\]

The difference \(\Delta_1(B|A,A')\) is the signal
amplitude when using a  small delay \(t\).

One now estimates this effect for the Doebner-Goldin equation in the original EPR state. 

The one-particle equation is given by (\ref{dg}) which one writes as
\[
i\hbar\partial_t\Psi_s =H_s\Psi_s=-\frac{\hbar^2}{2m_s}\Delta \Psi_s+iD_s\hbar(\Delta+ N)\Psi_s.
\]

For the two particle equation one takes 
\[i\hbar\partial_t\Psi_{ab} =H^{(1)}_a\Psi_ab+H^{(2)}_b\Psi_{ab}\]
without the \(Q\) term in (\ref{tpse}). 
The initial state \(\Phi\) is one of zero total momentum and one performs either
a momentum (\(A=p\)) or a position (\(A'=q\)) measurement on the first particle. One finds after some analysis (Svetlichny 2004d):
\[
\Delta_1(B|p,q) = 2 D_b\int\re(B\delta_w, (\Delta+N)\delta_w)\,d\mu(w)
\]
where \(\delta_w(y)= \delta(y-w)\) and \(\mu\)
a measure.  
 
Now \(N\) is  ill defined on \(\delta\) so
one uses a gaussian regularization 
\[\delta^{(s)}(y) = ({s \over \pi})^{n/2}e^{-s y^2}\]
with \(n\) the dimension of  space. As \(s\to \infty\) one has, as distributions,
\(\delta^{(s)}(y)=\delta(y)+O(s^{-1})\)

Asymptotic analysis now shows (Svetlichny 2004d)
 \[N(\delta^{(s)}) = 2n s \delta + \left({n \over 2}+1\right) \Delta \delta +
O(s^{-1}).\] 
 
Assume that \(B\delta_w\) is well defined and so \(B\delta^{(s)}_w =
B\delta_w +O(s^{-1})\), then in the end one finds
\[
\Delta_1(B|p,q) = 4sn D_b(\phi,B\phi) + O(1).
\]
Thus {\em even if the physical constant \(D_b\) is extremely
small, under extreme localization, (\(s\to\infty\)), the effect can be large\/}.
 
It is probably significant that not all nonlinear terms have this amplification effect.  Those that do not are,  the Bialynicki-Birula and Mycielski term \(p\ln|\Psi|\,\Psi\), the Kostin term \(iq \ln(\Psi/\bar\Psi)\,\Psi\), any  real nonlinear term added to the right-had side of the Doebner-Goldin equation, and any \(Q\) term in (\ref{tpse}).   Amplification seems to be a property of the precise diffeomorphism motivated nonlinearity.  

I thus answer my previous question:
\vskip 3pt
\noindent {\em At the Planck scale, nonlinear effects may be of the same order of magnitude as linear ones and still suffer large suppression at low energies\/}.
\vskip 3pt 
If such nonlinearities exist they would significantly alter our theories of physics at the Plank scale
and can offer a new alternative to current Planck-scale physics such as loop quantum gravity, M-theory, brane-world scenarios, quantum cosmology etc.

Though I downplay the prospect of solving some fundamental problem as a {\em  motivation\/} for nonlinear theories,  there is no harm in seeing what problems may be solvable once a non-liner theory is somehow introduced. At a first glance, Planck-scale nonlinearities could solve the following fundamental problems: cosmic homogeneity (space-like influences can homogenize as well as inflation), time's arrow (nonlinear quantum mechanics is generally time asymmetric), black hole information paradox (at Planck size, event horizons become permeable due to space-like influences), and possibly others.

To conclude I wish to present one final consideration that may make nonlinear quantum gravity plausible. Consider the familiar general relativistic dictum (apparently to to 
J.~A.~Wheeler):
\vskip 3pt
\noindent {\sl Matter tells space-time how to curve, space-time tells matter how to move.}
\vskip 3pt
Let me put a quantum ``spin"  on this: 
\vskip 3pt
\noindent {\sl Quantum matter tells quantum space-time how to be, quantum space-time tells quantum matter how to behave.}
\vskip 3pt
I've  changed ``curve" to ``be" and ``move" to ``behave" to accommodate my ignorance of what the appropriate quantum version should really be. 

I don't feel this is  a ``final" view. It's just the next step down from the present quantum-mechanics/general-relativity confrontation. On the level of the quantum dictum, the two are joined just barely, a dichotomy still exists (space and  matter), and it's up to better insights to go deeper.
 
Now quantum matter moves, in a first approximation, by a hamiltonian. By the quantum dictum, the hamiltonian must now depend on  quantum matter, this turns the quantum process nonlinear, as there is a back-reaction of matter on its own dynamics. One should {\em  metaphorically\/}  have:

\begin{eqnarray*}\label{schroedinger}
i\partial_t\Psi&=& H(\Psi), \\ \label{hamcurvature}
R(H)&=& 0.
\end{eqnarray*}
Here \(R\) is some operator (possibly differential) that \(H\) must satisfy. Linearity is \(D^2H=0\) where \(D\)  is the Fr\'echt derivative, and this must now be modified. The necessary existence of the second equation  has generally been ignored by investigators of nonlinear quantum mechanics. The possible use of projective Hilbert space (Kibble 1979, Leifer 1997)  as a scenario for quantum gravity is 
related to this, though I would classify these as willing approaches to nonlinearity and so I don't feel they are likely to unearth the true theory.

\subsection*{Acknowledgements} I wish to thank the organizers of the IQSA 2004 meeting for their invitation to give a talk. I thank the Conselho Nacional de Desenvolvimento Cient\'{\i}fico e Tecnol\'ogico (CNPq), and the Funda\c{c}\~ao Carlos Chagas Filho de Amparo \`a Pesquisa do Estado do Rio de Janeiro  (FAPERJ) for partial financial support. Some of the results here presented  were first obtained during  my sabbatical visit to the Mathematics Department of Rutgers University in 1992-3, for whose hospitality I am grateful. I also wish to thank Professor Gerald Goldin for interesting discussions on nonlinear quantum mechanics.

\section*{Bibliography}
\parskip 3pt
\noindent Abrams,~Daniel~S. and Lloyd~Seth (1998),
{\sl Physical Review Letters} {\bf 81} 3992.

\noindent Amelino-Camelia,~G. (2003), ``The three perspectives on the quantum-gravity problem and their implications for the fate of Lorentz symmetry", gr-qc/0309054.

\noindent Benatti,~F. and Floreanini,~R. (1996), {\sl Physics Letters} \textbf{B389}, 100.

\noindent  Benatti,~F. and Floreanini,~R.  (1999), {\sl Physics Letters} \textbf{B451}, 422.

\noindent Bialynicki-Birula,~I.  and Mycielski,~J.  (1976), {\sl Annals of Physics} {\bf 100} 62.

\noindent Bollinger,~J.~J. et al. (1989), {\sl Physical Review Letters} \textbf{63}, 1031.

\noindent Bona,~P. (1999), 
``Geometric Formulation of Nonlinear Quantum Mechanics for Density Matrices",
quant-ph/9910011.

\noindent Bugajski,~S.  (1991), {\em International Journal of
Theoretical Physics\/} {\bf 30} 961.

\noindent  Cassidy,~M.~J.  (1995), {\sl Physical Review} {\bf D52} 5676.

\noindent  Chupp,~T.~E. and Hoare,~R.~J.  (1990), {\sl Physical Review Letters} \textbf{64}, 2261.

\noindent Czachor,~M. (1999),
{\sl International Journal of Theoretical Physics} {\bf 38} 475.

\noindent Doebner,~H.-D. and Goldin,~G.~A. (1992), {\sl Physics Letters} {\bf A162} 397. 

\noindent Gheorghiu-Svirschevski,~S. (2002),
``A General Framework for Nonlinear Quantum Dynamics", quant-ph/0207042.

\noindent Gisin,~N. (1984), {\sl Physical Review Letters\/} {\bf 53} 1776.

\noindent Gisin,~N. (1989),  {\sl Helvetica Physica Acta \/} {\bf 62} 363.

\noindent Gisin,~N. (1990), {\sl Physics Letters A\/} {\bf 143} 1.

\noindent Goldin,~G.~A. and Svetlichny,~G. (1994),  {\sl Journal of Mathematical Physics} {\bf 35}, 3322.

\noindent Guz,~W. (1979), {\sl {\sl Reports on Mathematical Physics}} {\bf 16} 125.

\noindent Guz,~W. (1980),  {\sl Reports on Mathematical Physics} {\bf 17} 385.

\noindent Haag,~R. (1992), ``Local Quantum Physics", Springer Verlag, Berlin.

\noindent Haag,~R. and Bannier,~U. (1978), {\em Communications in Mathematical Physics\/}  {\bf 60} 1.

\noindent Hansson,~J. (2000),
``Nonlinear gauge interactions - A solution to the `measurement problem' in quantum  mechanics?",
quant-ph/0003083.

\noindent H\"ubsch~T. (1998), 
{\sl Modern Physics Letters} {\bf A13} 2503.

\noindent Kent,~A. (2002),
``Nonlinearity without Superluminality", quant-ph/0204106.

\noindent Kibble,~T.~W.~B. (1979), {\it Communications in Mathematical Physics} {\bf 65} 189.

\noindent Kostin,~M.~D. (1972), {\sl Journal of Chemical Physics\/} {\bf 57} 3589.

\noindent Leifer,~P. (1997), ``Nonlinear modification of quantum mechanics", hep-th/9702160.

\noindent Luecke,~W. (1999),
``Nonlocality in Nonlinear Quantum Mechanics",
quant-ph/9904016

\noindent Majumder,~P.~K. {\sl et al\/}. (1990), {\sl Physical Review Letters} \textbf{65}, 2931.

\noindent Markopoulou,~F. and Smolin,~L. (2003), ``Quantum Theory from Quantum Gravity", gr-qc/0311059.

\noindent Mavromatos,~N.~E. and Szabo,~R.~J. (2001), {\sl International Journal of Modern Physics}\ {\bf A16}, 209.

\noindent Meyer,~D.~A. (1996), ``Unitarity in one dimensional nonlinear quantum cellular automata", (quant-ph/9605023). 

\noindent Parwani,~R.~R. (2004),  ``An Information-Theoretic Link Between Spacetime Symmetries and Quantum Linearity", hep-th/0401190.

\noindent Polchinski,~J. (1991),  {\sl Physical Review Letters\/} {\bf 66} 397.

\noindent Simon,~C., Buzek,~V. and Gisin,~N.  (2001),
{\sl Physical Review Letters} {\bf 87}, 170405.

\noindent Singh,~T.~P. (2003),
``Quantum mechanics without spacetime III: a proposal for a non-linear Schrodinger equation", gr-qc/0306110.

\noindent Svetlichny,~G.  (1995),  {\sl Journal of Nonlinear Mathematical Physics} {\bf 2} 2.

\noindent Svetlichny,~G. (1997), ``On Relativistic Non-linear Quantum Mechanics" in  M. Shkil, A. Nikitin, and V. Boyko, eds., {\sl Proceedings of the Second International Conference ``Symmetry in Nonlinear Mathematical Physics. Memorial Prof. W. Fushchych Conference"}, Mathematics Institute, National Academy of Sciences of Ukraine, 1997, Vol. 2, p. 262. Proceedings are available online at the Mathematics Institute site: www.imath.kiev.ua/en/. 

\noindent Svetlichny,~G. (1998),
{\sl Foundation of Physics} {\bf 28} 131.

\noindent Svetlichny,~G. (2000),  {\sl Foundations of Physics} {\bf 30} 1819.

\noindent Svetlichny,~G. (2004a), 
{\sl Foundations of Physics Letters} {\bf 17} 197.

\noindent Svetlichny,~G. (2004b), {\sl Journal of Mathematical Physics} {\bf 45} 959.

\noindent Svetlichny,~G. (2004c) ``Informal Resource Letter -- Nonlinear quantum mechanics on arXiv up to August 2004", quant-ph/0410036.

\noindent Svetlichny,~G. (2004d) ``Amplification of acausal effects of nonlinear quantum mechanics under localization", quant-ph/0410186. 

\noindent Walsworth,~R.~L. et al. (1990), {\sl Physical Review Letters} \textbf{64}, 2599.

\noindent Weinberg,~S. (1989a), {\sl Physical Review Letters} {\bf 63} 485.

\noindent Weinberg,~S. (1989b), {\sl Annals of Physics (NY)} {\bf 194} 336.

\noindent Yurtsever,~U. and Hockney,~G. (2004) ``Signaling and the Black Hole Final State", hep-th/0402060.

\end{document}